\documentclass[preprint]{aastex}
\usepackage{natbib}
\input{epsf}
\usepackage{graphicx}
\usepackage{color}
\usepackage{yfonts}
\usepackage{pifont}
\usepackage{amsmath,amssymb}
\usepackage{multirow}
\usepackage{booktabs}
\usepackage{amsfonts}
\usepackage{subfigure}
\usepackage{geometry}
\usepackage{longtable}
\usepackage{lscape}
\usepackage{textcomp}
\usepackage[utf8]{inputenc}
\usepackage{rotating}

\usepackage{epstopdf}

\begin{document}
\title{TrendFolios\textsuperscript{\textregistered}: A Portfolio Construction Framework for Utilizing Momentum and Trend-Following In a Multi-Asset Portfolio
}
\author{Joseph Lu$^1$, Randall R. Rojas$^{2}$, Fiona C. Yeung$^3$, Patrick D. Convery$^{2}$}

\affil{$^1$Conscious Capital Advisors, LLC.}
\affil{$^2$Department of Economics\\University of
California, Los Angeles \\ Los
Angeles, CA 90095}
\affil{$^3$Walmart Labs}

\begin{abstract}
We design a portfolio construction framework and implement an active investment strategy utilizing momentum and trend-following signals across multiple asset classes and asset class risk factors. We quantify the performance of this strategy to demonstrate its ability to create excess returns above industry standard benchmarks, as well as manage volatility and drawdown risks over a 22+ year period. 

\end{abstract}

\section{Introduction}
 
Conscious Capital Advisors is an investment advisory firm that combines rigorous research with proprietary analytics to construct and manage portfolios with the aim of delivering superior risk-adjusted returns and long-term value for our clients. Our firm approach merges deep manager research expertise with innovative technologies, enabling efficient implementation; reducing both implicit and explicit costs. This paper examines one of our most basic algorithms, allowing readers to evaluate our investment philosophy and potential for generating long-term excess returns.

TrendFolios\textsuperscript{\textregistered}, like other momentum or trend-following approaches, aims to buy or hold assets with rising price trends and sell or not hold assets with price trends that are falling within its investment portfolio. These types of strategies analyze historical prices and returns across multiple timeframes to generate technical studies-such as moving averages, volatility measures, trading ranges, and price spreads-which are then converted into indicators or trading signals. Notably, momentum or trend-following strategies typically do not incorporate underlying economic theories or traditional valuation methods derived from corporate fundamentals. Liquidity and trading volume are also often critical factors in momentum and trend-following strategies, as they must be sufficient to accommodate time-sensitive execution when trading signals occur. Holding periods can range from minutes to weeks, with stop-loss orders often employed for drawdown management during volatile market environments.

Successful implementation of such strategies requires precise calibration of multiple variables: trend identification methodology, entry-exit point identification, position sizing, risk management, trade implementation, among others. The complexity of optimizing these interdependent variables has led to a large variety of approaches within these types of strategies on the market. TrendFolios\textsuperscript{\textregistered} sets forth a robust, streamlined methodology for addressing these challenges and demonstrates its effectiveness through empirical evidence.

\section{Literature Review}

Momentum and trend-following strategies have a well-documented history of performance across market cycles, with extensive academic research validating their effectiveness in delivering excess returns while managing volatility and limiting drawdowns. Early work by \cite{Jegadeesh1993} established the ``momentum effect,'' showing that ``momentum investing strategies provide abnormal returns in different stock markets,'' with profitability that was ``not due to their systematic risk or to delayed stock price reactions to common factors." \cite{Carhart1997} also famously observed that ``mutual funds with high returns last year have higher-than-average expected returns next year, but not in years thereafter.''

Since these seminal publishings, general consensus has concluded that the momentum effect has its long-term effectiveness rooted in the fundamental aspects of slow-changing human behavior. {\cite{Antonacci2015} explains that ``the most common explanations for both momentum and trend-following profits have to do with behavioral factors, such as anchoring, herding, and the disposition effect.'' This behavioral basis is further supported by \cite{Chen2014}, who identify that ``anchoring bias'' and the ``disposition effect''' are ``important factors explaining the momentum effect.'' Multiple literature reviews (\cite{Du2010}; \cite{Sondergaard2010}; \cite{Dhankar2016}) confirm that risk adjustments (i.e. beta, size, book-to-market) ``do not appear to explain momentum'' suggesting a non-rational, behavioral component.

The historical outperformance of momentum and trend-following strategies is well-documented in academic literature. \cite{Lemperiere2014} established ``the existence of anomalous excess returns based on trend-following strategies across four asset classes (commodities, currencies, stocks, bonds) and over very long time scales.'' They used for their studies ``...both futures time series, that exist since 1960, and spot time series that allow us to go back to 1800 on commodities and indices.'' The excess returns were statistically significant and ``very stable, both across time and asset classes''. Similarly, \cite{Hurst2017} constructed a ``...time-series momentum strategy all the way back to 1880 using historical data from a number of sources''. They found that ``time-series momentum has been consistently profitable throughout the past 137 years.'', and \cite{Asness2013}, demonstrated ``evidence on the return premia to value and momentum strategies globally across asset classes''. \cite{Antonacci2015} analyzed absolute momentum for stocks, bonds, and real assets and determined that ``absolute momentum can effectively identify regime change and add significant value as ... trend following overlay.'' He concluded that ``trend determination through absolute momentum can help one navigate downside risk, take advantage of regime persistence, and achieve higher risk-adjusted returns.''

However, the post-2008 era has introduced new challenges. While \cite{Faber2006} demonstrated a trend-following methodology that ``improved risk-adjusted returns across five asset classes while achieving relatively lower volatility and drawdown,'' and \cite{Moskowitz2011} identified ``significant time series momentum in equity, currency, commodity, and bond futures,'' suggesting that a ``diversified portfolio of time series momentum strategies across all asset classes delivers substantial abnormal returns,'' the landscape has shifted. \cite{Lo2016} argues that increased central bank intervention has fundamentally altered market dynamics, creating what he terms ``adaptive markets'' where traditional strategies must evolve. We hypothesize that drastic monetary and fiscal policies since the Global Financial Crisis of 2008 have effectively crowded out traditional active management approaches. In addition, the evolution of technology has also compressed trading timeframes and increased the speed of trend analysis. \cite{Chincarini2012} documents how technological advancement has enabled shorter-term strategies to effectively counter-trade longer-term momentum positions, requiring new approaches to traditional momentum implementation.

Despite this recent shift, industry research indicates that momentum and trend-following strategies remain valuable tools for institutional investors and registered investment advisors alike, primarily due to their transparency, instinctual promise of safety, and ease of interpretability. In response to this continued demand, we have created TrendFolios\textsuperscript{\textregistered}, a portfolio construction framework built on established research demonstrating the simplicity and effectiveness of combining trend-following and momentum trading signals with inverse volatility weighting.

\cite{Plessis2016} studied a basic directional momentum strategy which relies on the idea that ``an asset with a higher than average return will continue to have a higher return in the next period.'' They then implemented a risk management approach using a volatility weighting methodology which reduced positions when volatility was high, and increased investment when volatility was low. They found for ``empirical results for momentum strategies applied to US industry portfolios'' such volatility weighting is ``relevant for the improvement in Sharpe ratios''. This finding was reinforced by \cite{Shimizu2020}, whose inverse volatility weighting methodology significantly outperformed market capitalization weighted portfolios across multiple markets. \cite{Clare2012} demonstrated for commodity futures, that when combined with proper diversification, momentum and trend-following strategies can ``lead to portfolios which offer attractive risk adjusted returns'' and that  ``robust alpha survives after accounting for a wide variety of systemic risk''. According to the paper, ``the marginal impact of applying trend-following methods far outweighs momentum and risk parity adjustments in terms of risk-adjusted returns and limiting downside risk. Overall, this emphasizes the importance of trend-following as an investment strategy in the commodity futures context.'' Clare et al. (2013) also studied the performance of the 1,000 largest US stocks from 1968 through 2011 using a variety of indexing approaches (relative to a market cap weighted index approach) and determined that there are a wide variety of alternative indexing approaches that outperform a standard market cap weighting, which included an inverse volatility scheme. This verifies that this weighting approach is applicable in asset classes outside of commodities.

We implement the empirical insights provided by these authors into TrendFolios\textsuperscript{\textregistered}. Built upon this foundation of academic research,  TrendFolios\textsuperscript{\textregistered} delivers a sophisticated yet accessible investment solution for institutional investors and registered investment advisors alike.

\section{Methodology}

TrendFolios\textsuperscript{\textregistered} integrates momentum investing, trend-following methodologies, and inverse volatility weighting with periodic rebalancing. The process employs a fully quantitative, rules-based approach for portfolio construction and implementation. This framework allows for customized security selection based on investor preferences. 
 
To demonstrate the effectiveness of our framework, we have constructed sample asset class strategies that utilize factor exchange traded funds (``ETFs'') or ETFs representing alternative investment categories. These strategies aim to generate excess returns over benchmarks by capturing trending risk factors (or categories) in equities, fixed income, and alternatives while maintaining the volatility management and downside protection inherent in momentum and trend-following approaches. We evaluate this methodology using historical data, specifically prices of U.S. securities from 1997 through 2024.

\subsection{Security Selection and Data Sources}

Table 1 presents the ETFs selected for the equity, fixed income, and alternative investment strategies, along with their asset class subset description, and associated asset class risk factors. Details of the asset class risk factors are discussed in Appendix A. Asset class risk factors or categories used in the strategy are as follows:

\begin{enumerate}
\item Equity Factors
\begin{itemize}
\item Valuation
\item Market capitalization
\item Geographic exposure
\end{itemize}
\item Fixed Income Factors:
\begin{itemize}
\item Inflation protection
\item Duration
\item Credit quality
\item Currency exposure
\end{itemize}
\item ``Alternative Investment'' Factors:
\begin{itemize}
\item Private capital
\item Real estate
\item Various Commodity Exposures
\end{itemize}
\end{enumerate}

Our analysis utilizes in-sample returns data for the ETFs, which range from December 23, 1997 through December 27, 2023. As we accumulate additional returns data post-publication, we will be able to evaluate the out-of-sample performance of the model. Our analysis utilizes only ETF data that were available for trading during each historical period, rather than index data. The model's universe expands as new ETFs are introduced, creating a realistic representation of implementable strategies over time. This conservative approach likely understates potential performance, as a broader opportunity set would have provided additional investment breadth for return generation.

\begin{table}[ht]
{\small
\centering
\title{\textbf{List of ETFs included in our Portfolio.}}\\
\hspace{-15mm}
\begin{tabular}{lllll}
  \hline
ETF Proxy & Asset Class	&Asset Class Subset Description		& Risk Factors \\ \hline
IWD	&Equity		&Large U.S. Growth Stocks				& Growth\\
IWF	&Equity		&Large U.S. Value Stocks					& Value\\
IWN	&Equity		&Small U.S. Growth Stocks				& Capitalization, Growth\\
IWO	&Equity		&Small U.S. Value Stocks					& Capitalization, Value\\
EFA	&Equity		&Large Non-U.S. ``Developed Market'' Stocks	&Domicile\\
SCZ	&Equity		&Small Non-U.S. ``Developed Market'' Stocks  &Domicile, Capitalization\\
EEM	&Equity		&Non-U.S. "Emerging Market" Stocks		&Domicile\\
TIP	&Fixed Income	&U.S. Inflation-Protected Bonds	&Inflation\\
SHY	&Fixed Income					&U.S. Short Duration Bonds		&Short Duration\\
TLT	&Fixed Income				&U.S. Long Duration Bonds		&Long Duration\\
LQD	&Fixed Income			&U.S. Investment Grade Credit		&IG Credit Spreads\\
HYG	&Fixed Income				&U.S. High Yield Credit			&HY Credit Spreads\\
BWX	&Fixed Income		&Non-U.S. ``Developed Market'' Bonds	&Developed Market Currencies\\
EMB	&Fixed Income				&Non-U.S. ``Emerging Market" Bonds	&Emerging Market Currencies\\
BIZD	&Alternative		&Private Equity						&Private Capital\\
BKLN&Alternative			&Private Credit						&Private Capital\\
RWR&Alternative				&Real Estate						&Real Estate\\
USO	&Alternative						&Energy Commodities				&Commodity\\
DBA	&Alternative						&Agricultural Commodities			&Commodity\\
DBB	&Alternative					&Industrial Metal Commodities			&Commodity\\
GLD	&Alternative					&Precious Metal Commodities			&Commodity\\
  \hline
   \label{table:assets}
\end{tabular}
\caption{List of ETFs included in our Portfolio.}}
\end{table}

\subsection{Trading Signal Generation}

The trading signal employs two primary components:
\begin{enumerate}
\item Momentum Signal: Utilizes a line-based approach, reflects quantitative momentum approaches
\item Trend-Following Signal: Utilizes a curved-based approach, reflects technical trend-following approaches
\end{enumerate}

Our trading algorithm diversifies across two existing methodologies inspired by \cite{Faber2006} and \cite{Moskowitz2011}, and is based on a temporal integration of relative volatility (or tracking error), spreads and relative return signals derived from price ratios, each consisting of different calendar-based time frames, ranging from daily to annual. These signals are assigned to two ranking logics that capture the relative price momentum action and trend dynamics. In the final step, the momentum and trend-following signals are fused via a ``majority of vote'' algorithm, which then triggers inclusion (or exclusion if it does not meet our strict performance criteria) of the asset in the portfolio. 

As part of the inputs used, relative returns are computed for various frequencies to capture short-term and long-term dynamics. Therefore, for a given frequency $\nu$ (e.g., 1 day, 5 days, and so on), we compute them according to:

\begin{eqnarray}
R_{t} = \frac{P_t - P_{t-1}}{P_{t-1} }\times 100\%
\end{eqnarray} 

\noindent where:
\begin{itemize}
\item $R_t$ is the relative return at time $t$
\item $P_t$ is the adjusted closing price ratio at time $t$
\item $P_{t-1}$ is the adjusted closing price ratio at time $t-1$
\item $\nu$ represents the time frequency (e.g., daily, weekly, monthly)
\end{itemize}

\noindent The next metric computed is a compounded relative return, $CR_{t}$, computed according to:

\begin{eqnarray}
CR_{t} = CR_{t-1} \left(1 + \frac{R^{\nu}_{t}}{100} \right)
\end{eqnarray} 

\noindent where:

\begin{itemize}
\item $CR_t$ is the compounded relative return at time $t$
\item $CR_{t-1}$ is the compounded relative return at time $t-1$
\item $R_{t}^{\nu}$ is the relative return for frequency $\nu$ at time $t$
\end{itemize}

\noindent which is then used to to compute the series of daily normalized returns, $R^{\nu = 1}_{t}$ given by

\begin{eqnarray}
R^{1}_{t} = \frac{CR_t - CR_{t-1}}{CR_{t-1} }\times 100\%
\end{eqnarray} 

For $\nu \geq 5$, the normalized returns are computed according to:
\begin{eqnarray}
R^{\nu \geq 5}_{t} = \left[ \left(1 + \frac{R^{1}_{t}}{100} \right) \left(1 + \frac{R^{1}_{t-1}}{100} \right)\cdots \left(1 + \frac{R^{1}_{t-\nu+1}}{100} \right) \right] -1
\end{eqnarray} 

\noindent for different frequencies. The current frequencies align to calendar based metrics. 

Similar to our normalized returns metrics, we compute respective volatility metrics for each of the returns frequencies according to

\begin{eqnarray}
\sigma^{\nu}_{t} = \sqrt{  \frac{\sum_{i=0}^{n} (R^{\nu}_{t-i} - \overline{R}^{\nu})}{n-1} }
\end{eqnarray}

\noindent where:

\begin{itemize}
\item $\sigma_{t}^{\nu}$  is the relative volatility at time $t$ for frequency $\nu$
\item $R_{t-1}^{\nu}$ is the relative return at time $t-1$ for frequency $\nu$ 
\item $\overline{R}^{\nu}$ is the mean relative return for frequency $\nu$ 
\item $n$  is the number of periods in the calculation window
\end{itemize}

The spread signals, $S^{\nu}_{t}$ are computed based on the daily relative returns $R^{1}_{t}$ as follows:

\begin{eqnarray}
S^{\nu}_{t} = \frac{R^{1}_{t} - \overline{R}^{\nu}}{\overline{R}^{\nu}} + \frac{\sigma^{\nu}_{t}}{100}
\end{eqnarray}

\noindent where:

\begin{itemize}
\item $S_{t}^{\nu}$  is the spread at time $t$ for frequency $\nu$
\item $R_{t}^{1}$ is the daily relative return at time $t$ 
\item $\overline{R}^{\nu}$ is the mean relative return for frequency $\nu$ 
\item $\sigma_{t}^{\nu}$ is the relative volatility at time $t$ for frequency $\nu$
\end{itemize}

We then take all the calculation results for each item in our investment universe, and then assign a 1 if the investment meets the portfolio inclusion criterion, and a 0 otherwise. Across observation timeframes, this is represented as a binary matrix. The results from this step are integrated in both momentum and trend-following signals. A combination logic is used with the momentum and trend-following signals to finalize the inclusion/exclusion decision on each investment.

\subsection{Portfolio Construction and Trade Implementation}

After identifying all the positions that meet the inclusion criteria, we construct a portfolio with its components weighted using a standard inverse volatility method, but substitute standard deviation for tracking error (TE). Tracking error is the amount of volatility an asset has relative to a benchmark. The tracking error quantifies how closely the portfolio follows its benchmark by measuring the standard deviation of the difference between portfolio and benchmark returns:

\begin{eqnarray}
TE = \sqrt{ \frac{1}{n-1} \sum_{i=1}^{n} [(R_{p, i} - R_{b, i}) - (\overline{R}_p - \overline{R}_b )]^2   }
\end{eqnarray} 

\noindent where:

\begin{itemize}
\item $R_{p, i}$ is the portfolio return at time $i$
\item $R_{b, i}$ is the benchmark return at time i
\item $\overline{R}_p$ is the mean portfolio return over the period 
\item $\overline{R}_b$  is the mean benchmark return over the period
\item $n$ is the number of periods
\end{itemize}

Portfolio allocations are then determined using an inverse risk weighting approach - meaning we allocate more capital to less volatile factors and less capital to more volatile ones, as indicated by tracking error. For each asset i, the weight wi is calculated as:

\begin{eqnarray}
\omega_i = \frac{ \frac{1}{TE_i}}{\sum_{j=1}^{n} (1/TE_j)}
\end{eqnarray} 

\noindent where:

\begin{itemize}
\item is the relative volatility of asset $i$
\item $n$ is the total number of selected assets
\end{itemize}

With this weighting scheme, we construct a portfolio in which each asset in the portfolio contributes approximately the same amount of relative risk to the overall portfolio, and the overall tracking error of the portfolio can be lowered. Every two weeks, we analyze each risk factor's attractiveness relative to its benchmark, and then the portfolio is then methodically traded to reflect these new weightings every two weeks, ensuring our risk management remains structured and disciplined. 

\section{TrendFolios\textsuperscript{\textregistered} Asset Class Strategy Results}

We evaluate the effectiveness of the TrendFolios\textsuperscript{\textregistered} portfolio construction framework by comparing performance created by each asset class strategy against established industry standard benchmarks. For the performance of the equity strategy, we use the S\&P 500 index as our benchmark. For both fixed income and alternative investment strategies, the Bloomberg U.S. Aggregate Bond Index serves as the benchmark. The decision to benchmark alternatives against the bond index reflects the idea that most ``alternative'' investments generally serve as some form of fixed income substitute in portfolio allocation.

We also construct a moderate balanced portfolio that integrates all three asset class strategies, allocating 60\% to equities, 30\% to fixed income, and 10\% to alternatives. A static 10\% allocation to alternatives is taken from the fixed income allocation, representing again, alternatives as a fixed income substitute. We evaluate this strategy against the traditional 60/40 portfolio (60\% S\&P 500 Index, 40\% Bloomberg U.S. Aggregate Bond Index), which remains a standard allocation for U.S. investors with a moderate risk profile. 

To provide comprehensive context for capital allocators, we present both gross-of-fee and hypothetical net-of-fee performance metrics, enabling a clear assessment of the strategy's economic value proposition. The current fee assumption is 0.55\% of assets under management and is deducted monthly.

\subsection{TrendFolios\textsuperscript{\textregistered} Fixed Income Strategy}

The TrendFolios\textsuperscript{\textregistered} Fixed Income Strategy applies momentum and trend-following signals to a diversified universe of fixed income ETFs representing key risk factors including inflation protection, duration, credit quality, and currency exposures. The strategy utilizes a quantitative, rules-based approach that combines line-based momentum signals with curved-based trend-following indicators to identify attractive fixed income segments. Portfolio construction employs inverse volatility weighting, allocating more capital to less volatile fixed income factors and less capital to more volatile ones. The strategy conducts analysis every two weeks to determine each risk factor's attractiveness relative to its benchmark, with methodical portfolio rebalancing to reflect these new weightings, maintaining structured and disciplined risk management.

\begin{figure}[!ht]
\centering
\textbf{Rolling Annualized Excess Returns of theTrendFolios\textsuperscript{\textregistered}  Fixed Income Strategy}\\
\includegraphics[width=120mm]{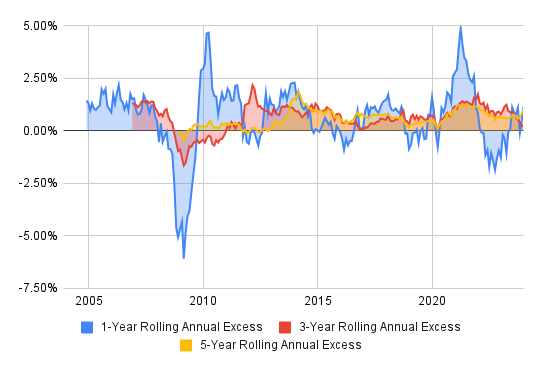}
\caption{1-Year (blue), 3-Year (red), and 5-Year (yellow) rolling annual excess returns from the TrendFolios\textsuperscript{\textregistered} Fixed Income Strategy backtest performance.}
\label{fig:trendfolio} 
\end{figure}

\begin{figure}[!ht]
\centering
\textbf{Growth of a Dollar (\$) in the TrendFolios\textsuperscript{\textregistered}  Fixed Income Strategy
}\\
\includegraphics[width=130mm]{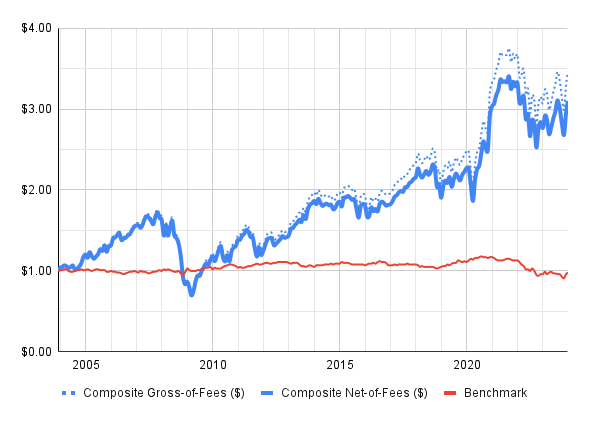}
\caption{The TrendFolios\textsuperscript{\textregistered} Fixed Income Strategy (blue) backtest performance against the Bloomberg U.S. Aggregate Bond Index (red).
}
\label{fig:trendfolio2} 
\end{figure}
\clearpage

\begin{table}[h!]
\centering
\textbf{TrendFolios\textregistered{} Fixed Income Strategy Annualized Performance}
\vspace{5pt} 
\begin{tabular}{lccccc}
\hline
\textbf{Annualized as of 12/26/2023} & \textbf{1-Year} & \textbf{3-Year} & \textbf{5-Year} & \textbf{10-Year} & \textbf{Since Inception} \\ \hline
Composite Net Return (\%)            & 12.09 & 0.73 & 10.22 & 5.22 & 5.79                   \\ 
Composite Gross Return (\%)          & 12.70 & 1.28 & 10.82 & 5.79 & 6.38                 \\ 
Index Return (\%)                    		& 2.08 & -5.74 & -1.37 & -0.69 & -0.10                 \\ 
Excess Return (net)                          & 10.01 & 6.46 & 11.59 & 5.90 & 5.89                   \\ 
Excess Return (gross)                	& 10.62 & 7.02 & 12.19 & 6.48 & 6.48                   \\ 
Composite Standard Deviation         & 18.59 & 17.41 & 19.03 & 15.82 & 17.49                  \\ 
Index Standard Deviation             &9.22 & 7.36 & 6.37 & 4.96 & 4.51              \\ 
Composite Sharpe Ratio               &0.65 & 0.04 & 0.54 & 0.33 & 0.33               \\ 
Index Sharpe Ratio                   & 0.23 & -0.78 & -0.22 & -0.14 & -0.02               \\ 
Tracking Error                       & 15.11 & 14.25 & 17.37 & 15.08 & 17.45                  \\ 
Information Ratio                    & 0.66 & 0.45 & 0.67 & 0.39 & 0.34                   \\ \hline
\end{tabular}
\caption{Quantitative performance metrics for the TrendFolios\textregistered{} Fixed Income Strategy compared to the baseline performance.}
\label{tab:performance_metrics}
\end{table}

%
%
%
%

\begin{table}[h]
    \centering
  \textbf{TrendFolios\textregistered{} Fixed Income Strategy Calendar Year Performance}\\
    \label{tab:trendfolios_performance}
    \begin{tabular}{cccccc}
        \toprule
        \textbf{Calendar Year} & \textbf{Strategy } & \textbf{Strategy } & \textbf{Benchmark} & \textbf{Excess Return } & \textbf{Excess Return } \\
          \textbf{} & \textbf{(Gross)} & \textbf{(Net)} & & \textbf{(Gross)} & \textbf{(Net)} \\
        \midrule
        2023  & 12.70\%  & 12.09\%  & 2.08\%   & 10.62\%  & 10.01\%  \\
        2022  & -16.35\% & -16.82\% & -15.04\% & -1.30\%  & -1.77\%  \\
        2021  & 10.21\%  & 9.61\%   & -3.42\%  & 13.63\%  & 13.03\%  \\
        2020  & 34.27\%  & 33.55\%  & 5.41\%   & 28.87\%  & 28.15\%  \\
        2019  & 19.81\%  & 19.16\%  & 5.71\%   & 14.09\%  & 13.44\%  \\
        2018  & -11.16\% & -11.65\% & -2.78\%  & -8.38\%  & -8.87\%  \\
        2017  & 18.88\%  & 18.23\%  & 0.93\%   & 17.94\%  & 17.30\%  \\
        2016  & 1.55\%   & 1.00\%   & 0.00\%   & 1.55\%   & 1.00\%   \\
        2015  & -2.03\%  & -2.57\%  & -1.83\%  & -0.20\%  & -0.73\%  \\
        2014  & 0.00\%   & -0.55\%  & 3.81\%   & -3.81\%  & -4.36\%  \\
        2013  & 30.46\%  & 29.76\%  & -4.55\%  & 35.01\%  & 34.31\%  \\
        2012  & 15.27\%  & 14.64\%  & 0.92\%   & 14.35\%  & 13.72\%  \\
        2011  & -9.03\%  & -9.53\%  & 4.81\%   & -13.84\% & -14.34\% \\
        2010  & 18.03\%  & 17.39\%  & 1.96\%   & 16.07\%  & 15.43\%  \\
        2009  & 37.08\%  & 36.34\%  & -0.97\%  & 38.05\%  & 37.32\%  \\
        2008  & -46.71\% & -47.02\% & 3.00\%   & -49.71\% & -50.02\% \\
        2007  & 5.03\%   & 4.46\%   & 1.01\%   & 4.02\%   & 3.45\%   \\
        2006  & 19.55\%  & 18.90\%  & -1.00\%  & 20.55\%  & 19.90\%  \\
        2005  & 9.92\%   & 9.32\%   & -0.99\%  & 10.91\%  & 10.31\%  \\
        2004  & 16.35\%  & 15.72\%  & 0.00\%   & 16.35\%  & 15.72\%  \\
        \bottomrule
    \end{tabular}
      \caption{Calendar Year Performance Metrics for the TrendFolios\textregistered{} Fixed Income Strategy compared to the baseline performance.}
\end{table}

\clearpage

\subsection{TrendFolios\textregistered{}  Equity Strategy}

The TrendFolios\textregistered{} Equity Strategy implements momentum and trend-following methodology across equity ETFs representing various risk factors including valuation (growth vs. value), market capitalization (large vs. small), and geographic exposure (U.S., developed international, and emerging markets). This systematic approach uses a proprietary algorithm that combines quantitative momentum signals with technical trend-following indicators to determine portfolio inclusion or exclusion. The strategy employs tracking error measurements rather than standard deviation for its inverse volatility weighting scheme, allocating more capital to factors with lower tracking error and less capital to those with higher tracking error. This construction method aims to ensure each equity factor contributes approximately equal relative risk to the overall portfolio while capturing trending risk factors within the equity universe.

\clearpage
\begin{figure}[!ht]
\centering
\textbf{Rolling Annualized Excess Returns of the TrendFolios\textregistered{} Equity Strategy}\\
\includegraphics[width=120mm]{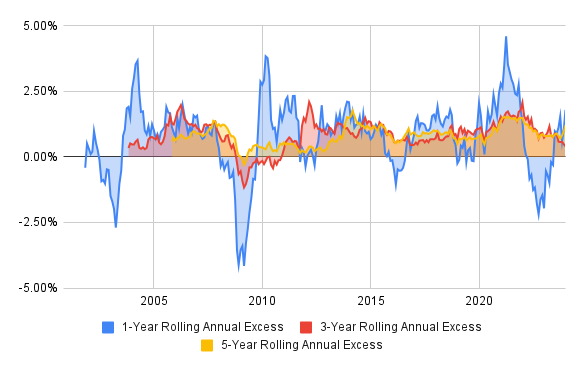}
\caption{1-Year (blue), 3-Year (red), and 5-Year (yellow) rolling annual excess returns from the TrendFolios\textregistered{}  Equity Strategy backtest performance.}
\label{fig:trendfolio} 
\end{figure}

\begin{figure}[!ht]
\centering
\textbf{Growth of a Dollar (\$) in the TrendFolios\textsuperscript{\textregistered} Equity Strategy}\\
\includegraphics[width=130mm]{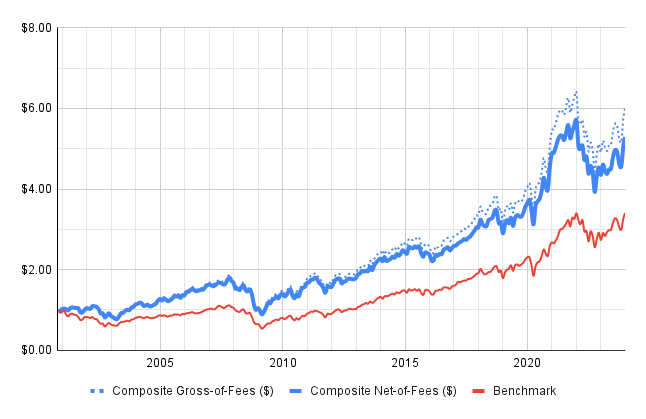}

\clearpage
\caption{The TrendFolios\textsuperscript{\textregistered}  Equity Strategy (blue) backtest performance against the S\&P 500 Index (red).}
\label{fig:trendfolio2} 
\end{figure}

\clearpage
\begin{table}[h!]
\centering
\textbf{TrendFolios\textregistered{} Equity Strategy Annualized Performance}
\vspace{5pt} 
\begin{tabular}{lccccc}
\hline
\textbf{Annualized (as of 12/26/2023)} & \textbf{1-Year} & \textbf{3-Year} & \textbf{5-Year} & \textbf{10-Year} & \textbf{Since Inception} \\ \hline
Composite Net Return (\%)   & 21.58  & 2.67   & 12.77  & 8.81  & 7.43   \\ 
Composite Gross Return (\%) & 22.24  & 3.24   & 13.39  & 9.41  & 8.02   \\ 
Index Return (\%)           & 24.37  & 8.35   & 13.74  & 9.91  & 5.75   \\
Excess Return (net)         & $-$2.80  & $-$5.67  & $-$0.96  & $-$1.10  & 1.67   \\ 
Excess Return (gross)       & $-$2.14  & $-$5.11  & $-$0.35  & $-$0.50  & 2.26   \\ 
Composite Standard Deviation & 16.41	&17.94	&19.39	&15.88	&16.36   \\ 
Index Standard Deviation     & 14.67	 &17.71	&18.64	&15.32	&15.42   \\ 
Composite Sharpe Ratio       & 1.32	&0.15	&0.66	&0.55	&0.45   \\ 
Index Sharpe Ratio           & 1.66	&0.47	&0.74	&0.65	&0.37   \\ 
Tracking Error              &6.08   & 7.76   & 7.18  &  5.96  &  6.26\\ 
Information Ratio            & -0.46  & -0.73  & -0.13 &  -0.18 &  0.27  \\ \hline
\end{tabular}
\caption{Quantitative performance metrics for the TrendFolios\textsuperscript{\textregistered} Equity Strategy compared to the baseline performance.}
\label{tab:performance_metrics}
\end{table}

\begin{table}[h]
    \centering
  \textbf{TrendFolios\textregistered{} Equity Strategy Calendar Year Performance}\\
    \label{tab:trendfolios_performance}
    \begin{tabular}{cccccc}
        \toprule
        \textbf{Calendar Year} & \textbf{Strategy } & \textbf{Strategy } & \textbf{Benchmark} & \textbf{Excess Return } & \textbf{Excess Return } \\
          \textbf{} & \textbf{(Gross)} & \textbf{(Net)} && \textbf{(Gross)} & \textbf{(Net)} \\
        \midrule
        2023  & 22.24\%  & 21.58\%  & 24.37\%  & -2.14\%  & -2.80\%  \\
        2022  & -23.20\% & -23.63\% & -19.49\% & -3.71\%  & -4.14\%  \\
        2021  & 17.21\%  & 16.58\%  & 27.01\%  & -9.80\%  & -10.44\% \\
        2020  & 34.65\%  & 33.93\%  & 16.13\%  & 18.52\%  & 17.80\%  \\
        2019  & 26.51\%  & 25.83\%  & 28.85\%  & -2.34\%  & -3.02\%  \\
        2018  & -4.29\%  & -4.82\%  & -6.36\%  & 2.07\%   & 1.54\%   \\
        2017  & 18.34\%  & 17.70\%  & 19.42\%  & -1.08\%  & -1.72\%  \\
        2016  & 9.80\%   & 9.20\%   & 9.62\%   & 0.17\%   & -0.42\%  \\
        2015  & -2.25\%  & -2.78\%  & -0.84\%  & -1.40\%  & -1.94\%  \\
        2014  & 7.87\%   & 7.28\%   & 11.23\%  & -3.36\%  & -3.95\%  \\
        2013  & 27.42\%  & 26.74\%  & 29.74\%  & -2.32\%  & -3.00\%  \\
        2012  & 15.30\%  & 14.68\%  & 13.43\%  & 1.88\%   & 1.25\%   \\
        2011  & -5.64\%  & -6.16\%  & -0.26\%  & -5.38\%  & -5.90\%  \\
        2010  & 19.63\%  & 18.98\%  & 12.79\%  & 6.83\%   & 6.18\%   \\
        2009  & 35.29\%  & 34.56\%  & 23.35\%  & 11.94\%  & 11.21\%  \\
        2008  & -37.26\% & -37.62\% & -38.32\% & 1.06\%   & 0.70\%   \\
        2007  & 3.47\%   & 2.91\%   & 3.27\%   & 0.21\%   & -0.36\%  \\
        2006  & 20.09\%  & 19.44\%  & 13.81\%  & 6.28\%   & 5.63\%   \\
        2005  & 8.08\%   & 7.49\%   & 3.05\%   & 5.03\%   & 4.44\%   \\
        2004  & 12.22\%  & 11.61\%  & 8.63\%   & 3.59\%   & 2.98\%   \\
        2003  & 34.47\%  & 33.75\%  & 26.09\%  & 8.38\%   & 7.66\%   \\
        2002  & -18.25\% & -18.70\% & -22.79\% & 4.54\%   & 4.08\%   \\
        2001  & 2.76\%   & 2.20\%   & -12.83\% & 15.59\%  & 15.02\%  \\
        \bottomrule
    \end{tabular}
      \caption{Calendar Year Performance Metrics for the TrendFolios\textregistered{} Equity Strategy compared to the baseline performance.}
\end{table}
\clearpage


\subsection{TrendFolios\textregistered{} Alternatives Strategy}

The TrendFolios\textregistered{} Alternatives Strategy applies the momentum and trend-following framework to alternative investment ETFs spanning multiple categories including private capital (both private equity and private credit), real estate, and various commodity exposures (energy, agricultural, industrial metals, and precious metals). The strategy serves as a fixed income substitute in portfolio allocation, utilizing the same dual-signal approach that combines momentum indicators with trend-following signals through a ``majority of vote'' algorithm. Like the other TrendFolios\textregistered{} strategies, it employs inverse volatility weighting for portfolio construction, but specifically targets alternative investments that historically could not be accessed through publicly traded exchanges. The strategy conducts bi-weekly analysis of each alternative category's attractiveness relative to fixed income benchmarks, methodically adjusting allocations to reflect these assessments.

\begin{figure}[!ht]
\centering
\textbf{Rolling Annualized Excess Returns of theTrendFolios\textsuperscript{\textregistered}  Alternatives Strategy}\\
\includegraphics[width=120mm]{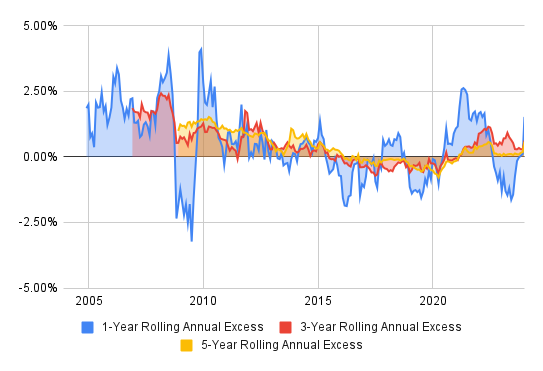} 
\caption{1-Year (blue), 3-Year (red), and 5-Year (yellow) rolling annual excess returns from the TrendFolios\textsuperscript{\textregistered} Alternatives Strategy backtest performance.}
\label{fig:trendfolio} 
\end{figure}

\begin{figure}[!ht]
\centering
\textbf{Growth of a Dollar (\$) in the TrendFolios\textsuperscript{\textregistered} Alternatives Strategy
}\\
\includegraphics[width=130mm]{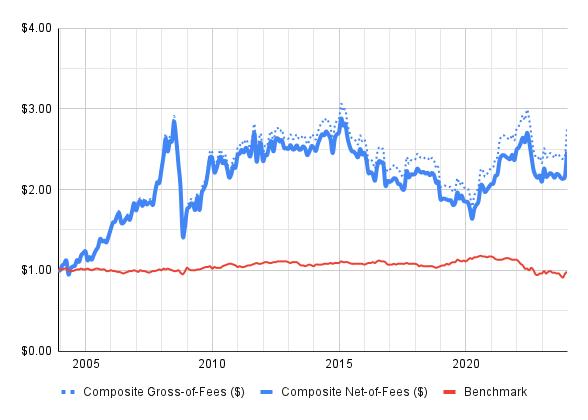}
\caption{The TrendFolios\textsuperscript{\textregistered} Alternatives Strategy (blue) backtest performance against the Bloomberg U.S. Aggregate Bond Index (red).
}
\label{fig:trendfolio2} 
\end{figure}
\clearpage

\begin{table}[h!]
\centering
\textbf{TrendFolios\textregistered{} Alternatives Strategy Annualized Performance}
\vspace{5pt} 
\begin{tabular}{lccccc}
\hline
\textbf{Annualized (as of 12/26/2023)} & \textbf{1-Year} & \textbf{3-Year} & \textbf{5-Year} & \textbf{10-Year} & \textbf{Since Inception} \\ 
\hline
Composite Net Return (\%)              & 7.22	&5.53	&4.45	&$-$0.73	&4.61 \\   
Composite Gross Return (\%)           &  7.81 & 6.11 & 5.02 & -0.18 & 5.19 \\ 
Index Return (\%)                    		&2.08 & -5.74 & -1.37 & -0.69 & -0.10 \\ 	
Excess Return (net)                  		&5.14 & 11.27 & 5.82 & -0.04 & 4.71 \\ 
Excess Return (gross)                 	  &  15.73 & 11.85 & 6.39 & 0.51 & 5.29  \\  
Composite Standard Deviation          & 10.08 & 9.59 & 10.32 & 11.00 & 15.80 \\ 
Index Standard Deviation               	& 9.22 & 7.36 & 6.37 & 4.96 & 4.51  \\ 
Composite Sharpe Ratio                	&   0.72 & 0.58 & 0.43 & -0.07 & 0.29  \\  
Index Sharpe Ratio                    	 &0.23 & -0.78 & -0.22 & -0.14 & -0.02 \\  
Tracking Error                         		 & 13.49 & 13.20 & 13.06 & 12.83 & 16.54 \\ 
Information Ratio           			 & 1.17 &0.90 & 0.53 & 0.04 & 0.28   \\ \hline
\end{tabular}
\caption{Quantitative performance metrics for theTrendFolios\textsuperscript{\textregistered} Alternatives Strategy compared to the baseline performance.}
\label{tab:annualized_performance}
\end{table}

\begin{table}[h]
    \centering
  \textbf{TrendFolios\textregistered{} Alternatives Strategy Calendar Year Performance}\\
    \label{tab:trendfolios_performance}
    \begin{tabular}{cccccc}
        \toprule
        \textbf{Calendar Year} & \textbf{Strategy } & \textbf{Strategy } & \textbf{Benchmark} & \textbf{Excess Return } & \textbf{Excess Return } \\
          \textbf{} & \textbf{(Gross)} & \textbf{(Net)} &  & \textbf{(Gross)} & \textbf{(Net)} \\
        \midrule
       2023  & 18.45\%  & 17.81\%  & 2.08\%   & 16.37\%  & 15.73\%  \\
        2022  & -15.27\% & -15.74\% & -15.04\% & -0.23\%  & -0.70\%  \\
        2021  & 21.15\%  & 20.49\%  & -3.42\%  & 24.56\%  & 23.91\%  \\
        2020  & 12.38\%  & 11.77\%  & 5.41\%   & 6.97\%   & 6.36\%   \\
        2019  & -1.46\%  & -2.00\%  & 5.71\%   & -7.18\%  & -7.72\%  \\
        2018  & -13.50\% & -13.98\% & -2.78\%  & -10.72\% & -11.20\% \\
        2017  & 5.33\%   & 4.76\%   & 0.93\%   & 4.40\%   & 3.82\%   \\
        2016  & -14.12\% & -14.60\% & 0.00\%   & -14.12\% & -14.60\% \\
        2015  & -9.03\%  & -9.53\%  & -1.83\%  & -7.19\%  & -7.70\%  \\
        2014  & 8.27\%   & 7.68\%   & 3.81\%   & 4.46\%   & 3.87\%   \\
        2013  & 1.14\%   & 0.59\%   & -4.55\%  & 5.69\%   & 5.13\%   \\
        2012  & 6.91\%   & 6.33\%   & 0.92\%   & 5.99\%   & 5.41\%   \\
        2011  & -2.38\%  & -2.92\%  & 4.81\%   & -7.19\%  & -7.73\%  \\
        2010  & 2.44\%   & 1.88\%   & 1.96\%   & 0.48\%   & -0.08\%  \\
        2009  & 37.43\%  & 36.69\%  & -0.97\%  & 38.40\%  & 37.67\%  \\
        2008  & -21.83\% & -22.28\% & 3.00\%   & -24.83\% & -25.28\% \\
        2007  & 29.38\%  & 28.68\%  & 1.01\%   & 28.37\%  & 27.67\%  \\
        2006  & 15.69\%  & 15.06\%  & -1.00\%  & 16.69\%  & 16.06\%  \\
        2005  & 23.39\%  & 22.72\%  & -0.99\%  & 24.38\%  & 23.71\%  \\
        2004  & 24.00\%  & 23.33\%  & 0.00\%   & 24.00\%  & 23.33\%  \\
        \bottomrule
    \end{tabular}
      \caption{Calendar Year Performance Metrics for the TrendFolios\textregistered{} Alternatives Strategy  compared to the baseline performance.}
\end{table}
\clearpage

%
%

\subsection{TrendFolios\textregistered{} Moderate Portfolio Strategy}

The TrendFolios\textregistered{} Moderate Portfolio Strategy integrates the three individual TrendFolios\textregistered{} strategies into a balanced portfolio framework, allocating 60\% to the Equity Strategy, 30\% to the Fixed Income Strategy, and 10\% to the Alternatives Strategy. This combined approach maintains the core momentum and trend-following methodology across multiple asset classes while providing a complete portfolio solution for investors with moderate risk tolerance. The 10\% allocation to alternatives comes from the fixed income portion, representing alternatives as a fixed income substitute. The integrated strategy preserves the individual components' disciplined portfolio construction methods, including quantitative trading signals, trend identification, and inverse volatility weighting. This unified framework allows for consistent application of momentum principles across diverse market segments within a single portfolio structure.

\begin{figure}[!ht]
\centering
\textbf{Rolling Annualized Excess Returns of theTrendFolios\textsuperscript{\textregistered}  Moderate Portfolio Strategy}\\
\includegraphics[width=120mm]{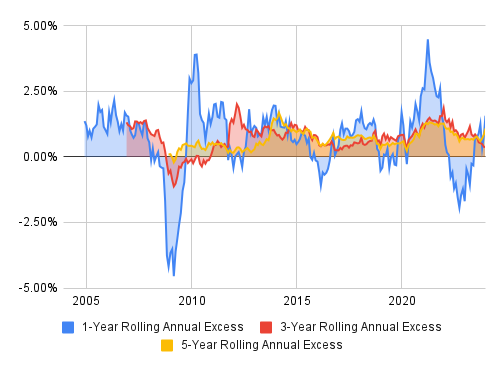} 
\caption{1-Year (blue), 3-Year (red), and 5-Year (yellow) rolling annual excess returns from the TrendFolios\textsuperscript{\textregistered}  Moderate Portfolio Strategy backtest performance.}
\label{fig:trendfolio} 
\end{figure}

\begin{figure}[!ht]
\centering
\textbf{Growth of a Dollar (\$) in the TrendFolios\textsuperscript{\textregistered} Moderate Portfolio
}\\
\includegraphics[width=130mm]{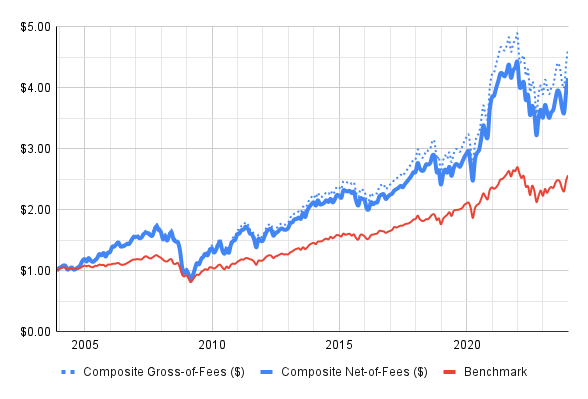}
\caption{The TrendFolios\textsuperscript{\textregistered}  Moderate PortfolioStrategy (blue) backtest performance against the Bloomberg U.S. Aggregate Bond Index (red).
}
\label{fig:trendfolio2} 
\end{figure}
\clearpage

\begin{table}[h!]
\centering
\textbf{TrendFolios\textregistered{} Moderate Portfolio Strategy Annualized Performance}
\vspace{5pt} 
\begin{tabular}{lccccc}
\hline
\textbf{Annualized (as of 12/26/2023)} & \textbf{1-Year} & \textbf{3-Year} & \textbf{5-Year} & \textbf{10-Year} & \textbf{Since Inception} \\ 
\hline
Composite Net Return (\%)              & 17.35 & 2.55 & 11.35 & 6.89 & 7.35 \\   
Composite Gross Return (\%)           &  17.98 & 3.11 & 11.95 & 7.48 & 7.94 \\ 
Index Return (\%)                    		&15.05 & 2.72 & 7.80 & 5.81 & 4.80 \\ 
Excess Return (net)                  		&2.29 & -0.17 & 3.54 & 1.08 & 2.55 \\ 
Excess Return (gross)                 	  & 2.93 & 0.40 & 4.15 & 1.66 & 3.14  \\  
Composite Standard Deviation          & 15.62 & 16.07 & 17.60 & 14.53 & 15.36 \\ 
Index Standard Deviation               	& 12.11 & 12.89 & 12.64 & 10.08 & 9.50  \\ 
Composite Sharpe Ratio                	&1.11 & 0.16 & 0.64 & 0.47 & 0.48 \\   
Index Sharpe Ratio                    	 &1.24 & 0.21 & 0.62 & 0.58 & 0.51 \\ 
Tracking Error                         		 &7.76 & 7.09 &7.77 & 6.61 &7.74\\
Information Ratio                       		& 0.44 &-0.01 & 0.47 & 0.17 & 0.33         \\ \hline
\end{tabular}
\caption{Quantitative performance metrics for theTrendFolios\textsuperscript{\textregistered} Moderate Portfolio Strategy compared to the baseline performance.}
\label{tab:annualized_performance}
\end{table}

\begin{table}[h]
    \centering
  \textbf{TrendFolios\textregistered{} Moderate Portfolio Strategy Calendar Year Performance}\\
    \label{tab:trendfolios_performance}
    \begin{tabular}{cccccc}
        \toprule
        \textbf{Calendar Year} & \textbf{Strategy } & \textbf{Strategy } & \textbf{Benchmark} & \textbf{Excess Return } & \textbf{Excess Return } \\
          \textbf{} & \textbf{(Gross)} & \textbf{(Net)} & & \textbf{(Gross)} & \textbf{(Net)} \\
        \midrule
    2023  & 19.13\%  & 18.49\%  & 15.05\%  & 4.08\%   & 3.43\%   \\
        2022  & -20.18\% & -20.63\% & -17.42\% & -2.76\%  & -3.20\%  \\
        2021  & 15.56\%  & 14.94\%  & 14.07\%  & 1.49\%   & 0.86\%   \\
        2020  & 32.39\%  & 31.68\%  & 12.53\%  & 19.86\%  & 19.15\%  \\
        2019  & 21.61\%  & 20.95\%  & 19.38\%  & 2.22\%   & 1.57\%   \\
        2018  & -7.28\%  & -7.80\%  & -4.65\%  & -2.63\%  & -3.14\%  \\
        2017  & 17.22\%  & 16.58\%  & 11.70\%  & 5.52\%   & 4.89\%   \\
        2016  & 4.81\%   & 4.23\%   & 5.85\%   & -1.04\%  & -1.61\%  \\
        2015  & -2.72\%  & -3.25\%  & -1.01\%  & -1.71\%  & -2.25\%  \\
        2014  & 5.59\%   & 5.01\%   & 8.30\%   & -2.71\%  & -3.29\%  \\
        2013  & 25.50\%  & 24.82\%  & 14.93\%  & 10.57\%  & 9.89\%   \\
        2012  & 14.49\%  & 13.87\%  & 8.40\%   & 6.08\%   & 5.46\%   \\
        2011  & -6.06\%  & -6.58\%  & 2.10\%   & -8.16\%  & -8.68\%  \\
        2010  & 17.50\%  & 16.87\%  & 8.91\%   & 8.60\%   & 7.96\%   \\
        2009  & 36.58\%  & 35.85\%  & 13.54\%  & 23.04\%  & 22.31\%  \\
        2008  & -38.38\% & -38.73\% & -23.70\% & -14.68\% & -15.03\% \\
        2007  & 6.40\%   & 5.82\%   & 2.51\%   & 3.89\%   & 3.31\%   \\
        2006  & 19.53\%  & 18.88\%  & 7.69\%   & 11.84\%  & 11.20\%  \\
        2005  & 10.19\%  & 9.59\%   & 1.51\%   & 8.68\%   & 8.08\%   \\
        2004  & 14.76\%  & 14.13\%  & 5.18\%   & 9.57\%   & 8.95\%   \\        \bottomrule
    \end{tabular}
      \caption{Calendar Year Performance Metrics for the TrendFolios\textregistered{} Moderate Portfolio Strategy compared to the baseline performance.}
\end{table}
\clearpage


\section{Performance Results Discussion}

The TrendFolios\textregistered{} framework demonstrates varied performance across asset classes from 1997 through 2023. The fixed income strategy delivered the most consistent outperformance, generating substantial excess returns of 6.48\% annually since inception over the Bloomberg U.S. Aggregate Bond Index, with particularly strong results during the post-2016 period. This strategy maintained positive rolling 5-year excess returns throughout most of the evaluation period, with only a brief interruption during the 2008 financial crisis.

The equity strategy showed mixed results, producing excess returns of 2.26\% annually since inception versus the S\&P 500 Index. While underperforming in the post-Global Financial Crisis period, it demonstrated downside protection during market stress events including the 2020 pandemic. The performance differential with the benchmark became more pronounced after 2020, suggesting improved effectiveness in recent market environments.

The alternatives strategy functioned effectively as a fixed income substitute, generating 5.29\% excess annual returns since inception compared to the Bloomberg U.S. Aggregate Bond Index. Its success stems from a diverse investment universe with varying correlation profiles, enabling dynamic tactical allocation across equity, fixed income, and commodity risk factors.

When combined into a moderate balanced portfolio (60\% equity, 30\% fixed income, 10\% alternatives), the approach delivered 3.14\% annual excess returns since inception versus a traditional 60/40 portfolio. Risk-adjusted metrics show information ratios ranging from 0.28 to 0.34 across strategies, indicating consistent generation of excess returns relative to the tracking error assumed. The methodology's inverse volatility weighting approach appears particularly effective in the fixed income space, with metrics showing controlled volatility relative to return generation.

The framework's performance across various market regimes validates the ongoing relevance of momentum and trend-following strategies, even amid unprecedented central bank intervention and technological advancement in market analysis.

\section{Conclusion}

TrendFolios\textregistered{} is an elegant solution for institutional investors and registered investment advisors to capitalize on the enduring value of systematic trend-following and momentum strategies. By combining established academic principles with practical implementation considerations, we have developed a framework that delivers meaningful results across multiple asset classes. The strategy's effectiveness is particularly evident in fixed income markets, where it has generated consistent excess returns through various market environments.

Our research validates the ongoing relevance of trend-following approaches, even in an era of unprecedented market intervention and technological advancement. While equity markets have presented unique challenges post-GFC, our multi-factor approach has proven robust, with particularly strong results in fixed income and alternatives. The framework's ability to adapt across different market regimes demonstrates the fundamental strength of combining momentum signals with sophisticated risk management techniques.
The alternatives component of TrendFolios\textregistered{} has proven especially valuable, functioning as an effective fixed income substitute while providing dynamic tactical allocation benefits. This success highlights the advantages of our systematic approach to capturing diverse risk premia across asset classes. The strategy's transparent, rules-based methodology offers institutional investors a practical solution for implementing academic insights in portfolio management.

Looking forward, we have identified multiple avenues for enhancing TrendFolios\textregistered{} effectiveness. These developments including parameter optimization, increased data frequency, advanced risk modeling, and universe expansion-build upon our proven foundation while maintaining the strategy's core benefits of transparency and efficiency. As markets continue to evolve, we believe our systematic approach to capturing trend and momentum effects, combined with sophisticated risk management, positions TrendFolios\textregistered{}  to deliver continued value for investors seeking robust, research-driven portfolio solutions.

\section{Future Research and Development Roadmap}

While impressive results have been generated from applying our TrendFolios\textregistered{}  framework on factor ETFs, we view the base algorithm as a foundation for continued innovation from our firm. We seek to continue our research with a focus in several key areas that we believe will greatly enhance the existing framework?s effectiveness and broaden its application:

\begin{enumerate}
\item {\bf Parameter Optimization-} The current model employs standardized calendar-based evaluation periods across all asset classes. We seek to develop asset-specific parameters that recognize the unique behavioral characteristics of each market segment, implementing adaptive time frame selection to improve signal quality and market regime identification
\item {\bf Increased Data Frequency-} Our current analysis utilizes only daily data, and we have the capacity to conduct analysis at intraday frequencies. We seek to expand our study to intraday analysis and develop the technological infrastructure necessary to capture shorter-term inefficiencies with more frequent trading.
\item {\bf Enhanced Drawdown Management-} Stop-loss mechanisms are a common downside protection strategy among systematic traders that has not been implemented in the current framework. We seek to explore various stop-loss mechanisms and dynamic risk controls to enhance risk management.
\item {\bf Increased Algorithm Sophistication-} Building on our current trading signals, we will incorporate more advanced algorithms in this area, which could utilize new artificial intelligence developments, as well as novel forecasting methodologies. These enhancements would aim to improve pattern recognition and adapt more effectively to evolving market conditions.
\item {\bf Enhanced Risk Forecasting-} Our inverse volatility approach can be enhanced through correlation forecasting and more dynamic risk parity implementation. This evolution might enable more responsive asset allocation while optimizing implementation efficiency.
\item {\bf Signal Fusion-} Consider weighting the frequency moments rather than using the current straight majority of vote algorithm when fusing signals.
\item {\bf Investment Universe Expansion-} We seek to significantly broaden our investment opportunity set, evaluating additional risk factors, fundamental indices, and additional alternative assets such as digital assets (cryptocurrencies), to capture diverse sources of return while maintaining liquidity and transparency.
\end{enumerate}

We believe TrendFolios\textregistered{} provides us with a strong start to demonstrate Conscious Capital Advisors' capacity for innovation in systematic investment management. Through ongoing research and development, we aim to enhance the model?s effectiveness across multiple dimensions, and aim to create a powerful investment solution to meet the investment challenges of the modern age.

\begin{appendix}
\section{Asset Class Definitions}

{\it Asset Class} An asset class is a group of investments that exhibit similar characteristics, behave similarly in the marketplace, and are subject to the same laws and regulations.

Key characteristics of an asset classes include:

\begin{itemize}
\item Assets within the class should be relatively homogeneous
\item Assets within the class should be diversifying
\item Asset classes should be mutually exclusive
\item The asset class should be large enough to be able to absorb a significant fraction of investors' investable wealth
\item The asset class should have the potential to earn positive real returns over time
\end{itemize}

Types of Asset Classes include:

\noindent {\bf Equities}

Equities represent partial ownership in companies, offering investors a way to participate in corporate growth and value creation over time. This ownership stake provides a claim on a company's earnings and assets, but by legal structure, equity investors are the lowest on the priority list to receive assets in the case of a bankruptcy, after creditors.

Key characteristics of equities include:

\begin{itemize}
\item Represent residual claims on company assets and cash flows
\item Have unlimited upside potential but limited downside (to zero)
\item Generally provide higher expected returns than fixed income over long periods
\item Typically demonstrate higher volatility than fixed income
\item May provide partial inflation protection through earnings growth
\item Can provide income through dividend payments
\item Have indefinite life unless company is liquidated or acquired
\item Highest long-term returns historically
\item Higher volatility
\item Suitable only for those with a long-term horizon
\item Factors include capitalization, valuation and domicile
\end{itemize}

Private equity is a type of equity investment but with differing legal risk which makes it a different asset class. Limited Partnership (LP) agreements expose investors to additional liabilities.

Non-U.S. equity is a type of equity which includes legal risk due to a philosophical difference about the role of capital or enterprise in society.  This legal risk can differ from country to country or culture to culture.

\noindent {\bf Key Equity Risk Factors:}

\begin{itemize}
\item {\it Market Capitalization} captures the size effect through free-float market value, with smaller companies historically delivering higher risk-adjusted returns versus large caps over a long investment horizon.
\item {\it Valuation} measures relative cheapness through metrics like P/E, P/B, and EV/EBITDA, identifying potential mispricings and mean reversion opportunities
\item {\it Domicile} factors reflect country-specific risks including political environment, regulatory frameworks, and market accessibility, particularly relevant when comparing developed vs emerging markets
\end{itemize}

\noindent {\bf Fixed Income}

Fixed income securities are debt instruments that provide regular income payments and return of principal at maturity. They represent lending and crediting an entity in search of capital, be it a commercial or government interest. Fixed income investors have priority over equity investors in a bankruptcy situation, and thus legally have a reduction of risk.

Key characteristics include:

\begin{itemize}
\item Contractual claims on issuer payments
\item Generally lower risk than equities due to priority in capital structure
\item Provide regular, predictable income streams
\item Have finite life (maturity date)
\item Returns primarily from interest payments rather than capital appreciation
\item Generally more sensitive to interest rate changes and credit risk
\item Limited upside potential compared to equities
\item Lower returns but more stable
\item Lower volatility
\item Generally liquid but varies by issue
\item Investment horizon varies based on maturity
\item Factors include duration, credit spread and currency.
\end{itemize}

Private credit is a type of fixed income investment with different legal risk which makes it a different asset class

Non-U.S. fixed income is a type of fixed income which includes legal risk due to a philosophical difference about the role of capital in society. Legal risk can differ from country to country or culture to culture.  Also, because of interest rate parity, non-U.S. fixed income risk can exist and is manifested as currency risk.

\noindent {\bf Fixed Income Risk Factors:}

\begin{itemize}
\item {\it Duration} quantifies interest rate sensitivity, measuring price changes for given yield moves. Longer duration typically commands higher term premium but increases rate risk
\item {\it Credit spread} captures excess yield over risk-free rates, compensating for default risk and reflecting corporate or sovereign issuer fundamentals. Spread changes drive returns beyond rate movements
\item {\it Currency} exposure impacts returns through exchange rate fluctuations and interest rate differentials between countries. Can be actively managed or hedged in portfolios depending on investment viewpoints and risk tolerance
\end{itemize}

\noindent {\bf Alternatives}

``Alternative'' investments are broad category investment offerings that historically could not be accessed through publicly traded exchanges, such as traditional stocks and bonds. 

Key characteristics of alternatives include:
\begin{itemize}
\item Wide variation in return potential
\item Wide variation in risk profile
\item Often illiquid
\item Often longer term investment horizon
\end{itemize}

Major categories of alternatives include:

\noindent {\bf Real Estate}

Represents ownership of land, and control of physical space. Considered a raw material. Human occupancy is what creates value. Subject to heavy legal and jurisdictional factors.

Key characteristics include:

\begin{itemize}
\item Provides potential for both income and capital appreciation
\item Often demonstrates low correlation with traditional assets
\item Can provide inflation protection through rent increases
\item Typically less liquid than traditional securities
\end{itemize}

\noindent {\bf Commodities} 

Commodities are traded on futures exchanges, which is another jurisdiction than stocks and contains different participants. ETFs and other legal changes are making this space more accessible. Commodities involve the ownership of raw materials and physical goods such as metals, oil, natural gas, and agricultural products traded in futures markets. Instead of providing capital to entrepreneurs, the investment is in the natural resources provided by the earth and prices are determined by supply and demand.

Key characteristics include:

\begin{itemize}
\item Wide variation in return potential
\item Wide variation in risk profile
\item Often illiquid
\item Often longer term investment horizon
\item Generally low correlation with traditional assets
\item Can be accessed through futures contracts or physical ownership
\item Typically no income generation, returns purely from price changes
\end{itemize}

\noindent {\bf Digital Assets} 

Represents another legal risk, as there is no well established jurisdiction.

Key characteristics include:


\noindent {\bf Private Capital and Hedge Funds} 

Private capital is agreement between two parties, it has lower government oversight than public capital and thus, enforcement potential in cases of bankruptcy.

Key characteristics include:

\begin{itemize}
\item Potentially higher returns but with increased risk
\item Longer investment horizons
\item Limited liquidity
\item Requires significant due diligence
\end{itemize}

\noindent {\bf Hedge Funds}

Ownership of proprietary investment or trading process. Hedge funds are essentially actively managed liquid portfolios deploying sophisticated strategies on tradable assets.

Key characteristics include:

\begin{itemize}
\item Aim for absolute returns regardless of market direction
\item Often use leverage and derivatives
\item May have low correlation with traditional assets
\item Usually very liquid, liquid alts
\item Generally higher fees than traditional investments
\item Manifestation of the human, of alpha
\end{itemize}

\begin{center}
{\bf Glossary of Performance Metrics}
\end{center}

{\it Information Ratio:} The information ratio (IR) measures portfolio returns beyond the returns of a benchmark, usually an index, compared to the volatility of those returns. The benchmark is typically an index representing the market or a particular sector or industry.

\begin{itemize}
\item The IR is often used to measure a portfolio manager's level of skill and ability to generate excess returns relative to a benchmark, but it also attempts to identify consistency of performance by incorporating a tracking error (TE) or standard deviation component into the calculation.
\item The IR goes beyond simple return metrics, offering a nuanced view of how consistently a portfolio outperforms its benchmark while accounting for how much risk is taken. This makes it a key tool for evaluating active investment strategies.
\end{itemize}

{\it Tracking Error:} Identifies the level of consistency in which a portfolio "tracks" the performance of an index. A lower tracking error indicates more consistent performance and less volatility relative to the benchmark, while a higher tracking error suggests greater variability and potential costs due to performance fluctuations.
\end{appendix}

{\it Annualized Return:} Represents the geometric mean of returns over multiple periods, converted to a yearly rate, showing how much an investment has grown on average per year.

{\it Standard Deviation:} Measures the dispersion of returns around their mean, indicating how volatile or risky an investment is - higher values suggest greater volatility and greater potential risk.

{\it Excess Return:} Is the difference between an investment's actual return and its benchmark return (like a market index), showing how much a portfolio outperformed or underperformed its reference point.

{\it Tracking Error:} Quantifies how closely a portfolio follows its benchmark by measuring the standard deviation of the difference between the portfolio and benchmark returns over time.

{\it Sharpe Ratio:} Measures how much excess return you receive for the additional volatility of holding an asset that could generate returns higher than a risk-free rate. A higher Sharpe ratio indicates better risk-adjusted performance. We modify our Sharpe Ratio by ignoring the risk free rate, essentially making a pure return to risk metric.

I{\it nformation Ratio:} Measures a portfolio manager's consistency in generating excess returns by dividing the average excess return by the tracking error, indicating skill in outperforming the benchmark.

{\it Sortino Ratio:} Only considers downside volatility, making it particularly useful for evaluating investments where upside volatility is not considered a risk. Like our modified Sharpe ratio, we modify the Sortino Ratio by not considering the risk free rate to create a pure return to risk ratio.

{\it Calmar Ratio:} Divides the average annual compound rate of return by the maximum drawdown risk, helping investors evaluate the relationship between return and worst-case risk over a specified time period.

\end{document}